\documentclass{birkjour}
 \newtheorem{thm}{Theorem}[section]

 \theoremstyle{definition}
 
 \theoremstyle{remark}
 
 \newtheorem*{ex}{Example}
 \numberwithin{equation}{section}
\usepackage[english]{babel}

\def\diag{{\rm diag}}

\def\R{{\mathbb R}}
\def\C{{\mathbb C}}
\def\H{{\mathbb H}}

\def\Adj{{\rm Adj}}
\def\cen{{\rm cen}}

\def\la{\langle}
\def\ra{\rangle}

\def\va{\triangle}

\def\cl{{C}\!\ell}

\def\Det{{\rm Det}}

\def\mod{{\,\rm mod}}
\allowdisplaybreaks

\begin{document}

%
%
%
%
%
%
%
%
%

\title[Basis-free Solution to Sylvester Equation in Clifford Algebra]
 {Basis-free Solution to Sylvester Equation in Clifford Algebra of Arbitrary Dimension}

\author[Dmitry Shirokov]{Dmitry Shirokov}

\address{%
HSE University\\
Myasnitskaya str. 20\\
101000 Moscow\\
Russia\\
{}\\
and\\
{}\\
Institute for Information Transmission Problems of Russian Academy of Sciences\\
Bolshoy Karetny per. 19\\
127051 Moscow\\
Russia}
\email{dm.shirokov@gmail.com}




\date{January, 2021}

\begin{abstract}
The Sylvester equation and its particular case, the Lyapunov equation, are widely used in image processing, control theory, stability analysis, signal processing, model reduction, and many more. We present basis-free solution to the Sylvester equation in Clifford (geometric) algebra of arbitrary dimension. The basis-free solutions involve only the operations of Clifford (geometric) product, summation, and the operations of conjugation. To obtain the results, we use the concepts of characteristic polynomial, determinant, adjugate, and inverse in Clifford algebras. For the first time, we give alternative formulas for the basis-free solution to the Sylvester equation in the case $n=4$, the proofs for the case $n=5$ and the case of arbitrary dimension $n$. The results can be used in symbolic computation.
\end{abstract}

\maketitle

\section{Introduction}

This paper is an extended version of the short note in Conference Proceedings \cite{ShirokovSylv}. We present for the first time the alternative formulas for the basis-free solution to the Sylvester equation in the case $n=4$ (see the remarks after Theorem \ref{thSylv4}), the proofs of Theorems \ref{thSylv5} and \ref{thgen}, and the simplification of the statement of Theorem \ref{thgen} in the case of odd $n=p+q$ (see the remarks after Theorem \ref{thgen}).

The Sylvester equation \cite{Sylv} is a linear equation of the form $AX-XB=C$ for known $A, B, C$ (quaternions, matrices, or multivectors depending on the formalism) and unknown $X$. The Sylvester equation and its particular case, the Lyapunov equation (with $B=-A^H$), are widely used in different applications -- image processing, control theory, stability analysis, signal processing, model reduction, and many more. In this paper, we study the Sylvester equation in Clifford's geometric algebra $\cl_{p,q}$ and present basis-free solution to this equation in the case of arbitrary $n=p+q$.

The Sylvester equation over quaternions corresponds to the Sylvester equation in geometric algebra of a vector space of dimension $n=2$, because we have the isomorphism $\cl_{0,2}\cong \H$. Thus the basis-free solution to the Sylvester equation in $\cl_{p,q}$, $p+q=2$, is constructed similarly to the basis-free solution to the Sylvester equation over quaternions. The same ideas as in the case $n=2$ work in the case $n=3$. The cases $n\leq 3$ are also discussed by Acus and Dargys~\cite{acusSylv}.

In this paper, we present basis-free solutions in the cases $n=4$ and $n=5$, which are the most important cases for the applications. The geometric algebra $\cl_{1,3}$ (the space-time algebra \cite{HestenesSTA}) of a space of dimension $4$ is widely used in physics. The conformal geometric algebra $\cl_{4,1}$ of a space of dimension $5$ is widely used in geometry, robotics, and computer vision (see \cite{bayro, dorst, hildenbrand, Hong, HitzerConf}). Also we present recursive basis-free formulas to the Sylvester equation in $\cl_{p,q}$ in the case of arbitrary $n=p+q$. They can be used in symbolic computation. We use our previous results on explicit and recursive formulas for the characteristic polynomial coefficients and inverse in Clifford algebras \cite{det}. Note also the papers on the characteristic polynomial \cite{Helm} and inverse \cite{dadbeh, rudn, acus, hitzer1, hitzer2}.

An arbitrary linear quaternion equation with two terms
\begin{eqnarray}
KXL+MXN=P\label{lqe}
\end{eqnarray} for known $K, L, M, N, P\in\H$ and unknown $X\in\H$ can be reduced to the Sylvester equation. Any nonzero quaternion $Q=a+bi+cj+dk\neq 0$, where $a, b, c , d\in\R$ are real numbers, and $i$, $j$, and $k$ are the quaternion units, is invertible and the inverse is equal to $$Q^{-1}=\frac{\overline{Q}}{Q \overline{Q}},$$
where $\overline{Q}:=a-bi-cj-dk$ is the conjugate of $Q$. Multiplying both sides of (\ref{lqe}) on the left by $M^{-1}$ and on the right by $L^{-1}$, we obtain $M^{-1}KX+XNL^{-1}=M^{-1}P L^{-1}$. Denoting
$A:=M^{-1}K$, $B:=-N L^{-1}$, $C:=M^{-1}P L^{-1}$,
we get the Sylvester equation
\begin{eqnarray}
AX-XB=C\label{Syl}
\end{eqnarray}
for known $A, B, C\in\H$ and unknown $X\in\H$ (see also \cite{1,Hongbo}).

Multiplying both sides of (\ref{Syl}) on the right by $-\overline{B}$, we get
\begin{eqnarray}
-AX\overline{B}+X B \overline{B}=-C \overline{B}.\label{1}
\end{eqnarray}
Multiplying both sides of (\ref{Syl}) on the left by $A$, we get
\begin{eqnarray}
A^2 X-AXB=AC.\label{2}
\end{eqnarray}
Summing (\ref{1}) and (\ref{2}) and using $B+\overline{B}\in\R$, $B\overline{B}\in\R$, we obtain
$$A^2 X-(B+\overline{B})A X+B \overline{B} X=AC-C\overline{B}.$$
If
$$D:=A^2-BA-\overline{B}A+B\overline{B}\neq 0,$$
then $D$ is invertible and we get the basis-free solution to (\ref{Syl}):
$$X=D^{-1}(AC-C\overline{B})=\frac{\overline{D}(AC-C\overline{B})}{D \overline{D}}.$$

\section{The Cases $n\leq 3$}

Let us consider the Clifford's geometric algebra $\cl_{p,q}$, $p+q=n$, \cite{Hestenes,Lasenby,Lounesto,LM,Port,Bulg} with the identity element $e$ and the generators $e_a$, $a=1, \ldots, n$, satisfying
$$e_a e_b+e_b e_a=2\eta_{ab}e,\qquad a, b=1, \ldots, n,$$
where $\eta=(\eta_{ab})=\diag(1,\ldots, 1, -1, \ldots, -1)$ is the diagonal matrix with its first $p$ entries equal to $1$ and the last $q$ entries equal to $-1$ on the diagonal. We call the subspace of $\cl_{p,q}$ of geometric algebra elements, which are linear combinations of basis elements with multi-indices of length $k$, the subspace of grade $k$ and denote it by $\cl^k_{p,q}$, $k=0, 1, \ldots, n$. We identify elements of the subspace of grade $0$ with scalars: $\cl^0_{p,q}\equiv \R$, $e\equiv 1$. Denote the operation of projection onto the subspace $\cl^k_{p,q}$ by $\la \quad \ra_k$. The center of $\cl_{p,q}$ is $\cen(\cl_{p,q})=\cl^0_{p,q}$ in the case of even $n$ and $\cen(\cl_{p,q})=\cl^0_{p,q}\oplus\cl^n_{p,q}$ in the case of odd $n$.

We use the following two standard operations of conjugation in $\cl_{p,q}$: the grade involution $\widehat{\quad}$ and the reversion (an anti-involution) $\widetilde{\quad}$
\begin{eqnarray}
&&\widehat{U}=\sum_{k=0}^n (-1)^k \la U \ra_k,\qquad \widehat{UV}=\widehat{U}\widehat{V},\qquad\forall U, V\in\cl_{p,q},\label{hat}\\
&&\widetilde{U}=\sum_{k=0}^n(-1)^{\frac{k(k-1)}{2}} \la U\ra_k,\qquad \widetilde{UV}=\widetilde{V}\widetilde{U},\qquad\forall U, V\in\cl_{p,q}.\label{tilde}
\end{eqnarray}

Let us consider the Sylvester equation in geometric algebra
\begin{eqnarray}
AX-XB=C\label{Sylv123}
\end{eqnarray}
for known $A, B, C\in\cl_{p,q}$ and unknown $X\in\cl_{p,q}$.

In the case $n=1$, the geometric algebra $\cl_{p,q}$ is commutative and we get $(A-B)X=C$. Denoting $D:=A-B$ and using\footnote{The definitions of adjugate $\Adj(D)$, determinant $\Det(D)$, and inverse $D^{-1}$ in $\cl_{p,q}$ for an arbitrary $n$ are given in \cite{det}.}
$$\Adj(D)=\widehat{D},\qquad \Det(D)=D\widehat{D}\in \cl^0_{p,q}\equiv\R,\qquad D^{-1}=\frac{\Adj(D)}{\Det(D)},$$
we conclude that if
\begin{eqnarray}
Q:=D\widehat{D}\neq 0,\label{Q1}
\end{eqnarray}
then
$$X=\frac{\widehat{D}C}{Q}.$$

In the case $n=2$, we can do the same as for the Sylvester equation over quaternions (see Introduction). Multiplying both sides of (\ref{Sylv123}) on the right by $-\widehat{\widetilde{B}}$ and on the left by $A$, we get
$$-AX\widehat{\widetilde{B}}+XB \widehat{\widetilde{B}}=-C \widehat{\widetilde{B}},\qquad A^2 X-AXB=AC.$$
Summing and using $\Det(B)=B \widehat{\widetilde{B}}\in\cl^0_{p,q}\equiv \R$, $B+\widehat{\widetilde{B}}\in\cl^0_{p,q}\equiv \R$, we get
\begin{eqnarray}
(A^2-(B+\widehat{\widetilde{B}})A+B\widehat{\widetilde{B}})X=AC-C \widehat{\widetilde{B}}.\label{g2}
\end{eqnarray}
Using
$$\Adj(D)=\widehat{\widetilde{D}},\qquad \Det(D)=D\widehat{\widetilde{D}}\in\cl^0_{p,q}\equiv \R,\qquad  D^{-1}=\frac{\Adj(D)}{\Det(D)},$$
we conclude that if
\begin{eqnarray}
Q:=D \widehat{\widetilde{D}}\neq 0\label{Q2},
\end{eqnarray}
then for $D:=A^2-(B+\widehat{\widetilde{B}})A+B\widehat{\widetilde{B}}$, we get
$$X=\frac{\widehat{\widetilde{D}}(AC-C \widehat{\widetilde{B}})}{Q}.$$

In the case $n=3$, we have $B \widehat{\widetilde{B}}\in\cl^0_{p,q}\oplus\cl^3_{p,q}=\cen(\cl_{p,q})$ and $B+\widehat{\widetilde{B}}\in\cl^0_{p,q}\oplus\cl^3_{p,q}=\cen(\cl_{p,q})$ and obtain again (\ref{g2}).
Using
$$\Adj(D)=\widehat{D} \widetilde{D} \widehat{\widetilde{D}},\qquad \Det(D)=D\widehat{D} \widetilde{D} \widehat{\widetilde{D}}\in\cl^0_{p,q}\equiv \R,\qquad  D^{-1}=\frac{\Adj(D)}{\Det(D)}$$
for $D:=A^2-(B+\widehat{\widetilde{B}})A+B\widehat{\widetilde{B}}$, we conclude that if
\begin{eqnarray}
Q:=D \widehat{D} \widetilde{D} \widehat{\widetilde{D}}\neq 0,\label{Q3}
\end{eqnarray}
then
\begin{eqnarray}
X=\frac{\widehat{D}\widetilde{D}\widehat{\widetilde{D}}(AC-C \widehat{\widetilde{B}})}{Q}.
\end{eqnarray}

\section{The Case $n=4$}

Let us consider one additional operation of conjugation $\va$ (compare with the grade involution (\ref{hat}) and the reversion (\ref{tilde}), see also \cite{det})
\begin{eqnarray}
U^\va&=&\sum_{k=0}^n (-1)^{\frac{k(k-1)(k-2)(k-3)}{4!}} \la U \ra_k\nonumber\\
&=&\!\!\!\!\!\sum_{k=0,1, 2, 3\mod 8}\la U\ra_k-\sum_{k=4, 5, 6, 7\mod 8}\la U\ra_k,\qquad \forall U\in\cl_{p,q}.\label{tri}
\end{eqnarray}
In the general case, we have $(UV)^\va\neq U^\va V^\va$ and $(UV)^\va\neq V^\va U^\va$.

\begin{thm}\label{thSylv4} Let us consider the Sylvester equation in $\cl_{p,q}$, $p+q=4$
\begin{eqnarray}
AX-XB=C,\label{Sylv4}
\end{eqnarray}
for known $A, B, C\in\cl_{p,q}$ and unknown $X\in\cl_{p,q}$.

If
\begin{eqnarray}
Q:=D\widehat{\widetilde{D}}(\widehat{D} \widetilde{D})^{\va}\neq 0,\label{Q4}
\end{eqnarray}
then
\begin{eqnarray}
X=\frac{\widehat{\widetilde{D}}(\widehat{D} \widetilde{D})^{\va} F}{Q},\label{X}
\end{eqnarray}
where
\begin{eqnarray}
D&:=&A^4-A^3(B+\widehat{\widetilde{B}}+\widehat{B}^\va+\widetilde{B}^\va)\label{D}\\
&+&A^2(B\widehat{\widetilde{B}}+B\widehat{B}^\va+B\widetilde{B}^\va+\widehat{\widetilde{B}}\widehat{B}^\va+ \widehat{\widetilde{B}}\widetilde{B}^\va+(\widehat{B}\widetilde{B})^\va)\nonumber\\
&-&A(B \widehat{\widetilde{B}} \widehat{B}^\va+ B \widehat{\widetilde{B}} \widetilde{B}^\va+ B(\widehat{B} \widetilde{B})^\va+ \widehat{\widetilde{B}} (\widehat{B} \widetilde{B})^\va)+B\widehat{\widetilde{B}}(\widehat{B} \widetilde{B})^{\va},\nonumber\\
F&:=&A^3 C-A^2C(\widehat{\widetilde{B}}+\widehat{B}^\va+\widetilde{B}^\va)\label{F}\\
&+&AC(\widehat{\widetilde{B}} \widehat{B}^\va+ \widehat{\widetilde{B}} \widetilde{B}^\va+ (\widehat{B} \widetilde{B})^\va)-C\widehat{\widetilde{B}}(\widehat{B} \widetilde{B})^{\va}.\nonumber
\end{eqnarray}
\end{thm}
As one of the anonymous reviewers of this paper noted, the formulas (\ref{Q4}), (\ref{X}), (\ref{D}), and (\ref{F}) can be rewritten using the new operation 
\begin{eqnarray}
B^\natural:=(\widehat{B}\widetilde{B})^\va\label{natural}
\end{eqnarray}
in the following form
\begin{eqnarray*}
&&Q=D\widehat{\widetilde{D}}D^\natural,\qquad X=\frac{\widehat{\widetilde{D}}D^\natural F}{Q},\\
&&D=A^4-A^3(B+\widehat{\widetilde{B}}+\widehat{B}^\va+\widetilde{B}^\va)+A^2(B\widehat{\widetilde{B}}+ B\widehat{B}^\va+B\widetilde{B}^\va+\widehat{\widetilde{B}}\widehat{B}^\va\\
&&+ \widehat{\widetilde{B}}\widetilde{B}^\va+B^\natural)-A(B \widehat{\widetilde{B}} \widehat{B}^\va+ B \widehat{\widetilde{B}} \widetilde{B}^\va+ BB^\natural+ \widehat{\widetilde{B}} B^\natural)+B\widehat{\widetilde{B}}B^\natural,\nonumber\\
&&F=A^3 C-A^2C(\widehat{\widetilde{B}}+\widehat{B}^\va+\widetilde{B}^\va)+AC(\widehat{\widetilde{B}} \widehat{B}^\va+ \widehat{\widetilde{B}} \widetilde{B}^\va+ B^\natural)-C\widehat{\widetilde{B}}B^\natural.\nonumber
\end{eqnarray*}
\begin{proof}
Multiplying both sides of (\ref{Sylv4}) on the right by $-\widehat{\widetilde{B}}(\widehat{B} \widetilde{B})^{\va}$, we get
\begin{eqnarray}
-AX\widehat{\widetilde{B}}(\widehat{B} \widetilde{B})^{\va}+X B\widehat{\widetilde{B}}(\widehat{B} \widetilde{B})^{\va}=-C\widehat{\widetilde{B}}(\widehat{B} \widetilde{B})^{\va}.\label{q1}
\end{eqnarray}
Multiplying both sides of (\ref{Sylv4}) on the right by $\widehat{\widetilde{B}} \widehat{B}^\va+ \widehat{\widetilde{B}} \widetilde{B}^\va+ (\widehat{B} \widetilde{B})^\va$  and on the left by $A$, we get
\begin{eqnarray}
&&A^2 X (\widehat{\widetilde{B}} \widehat{B}^\va+ \widehat{\widetilde{B}} \widetilde{B}^\va+ (\widehat{B} \widetilde{B})^\va) -AXB(\widehat{\widetilde{B}} \widehat{B}^\va+ \widehat{\widetilde{B}} \widetilde{B}^\va+ (\widehat{B} \widetilde{B})^\va)\nonumber\\
&&=AC(\widehat{\widetilde{B}} \widehat{B}^\va+ \widehat{\widetilde{B}} \widetilde{B}^\va+ (\widehat{B} \widetilde{B})^\va).\label{q2}
\end{eqnarray}
Multiplying both sides of (\ref{Sylv4}) on the right by $-(\widehat{\widetilde{B}}+\widehat{B}^\va+\widetilde{B}^\va)$ and on the left by $A^2$, we get
\begin{eqnarray}
&&-A^3 X(\widehat{\widetilde{B}}+\widehat{B}^\va+\widetilde{B}^\va)+ A^2 XB(\widehat{\widetilde{B}}+\widehat{B}^\va+\widetilde{B}^\va)\nonumber\\
&&= -A^2C(\widehat{\widetilde{B}}+\widehat{B}^\va+\widetilde{B}^\va).\label{q3}
\end{eqnarray}
Multiplying both sides of (\ref{Sylv4}) on the left by $A^3$, we get
\begin{eqnarray}
A^4X-A^3 XB=A^3C.\label{q4}
\end{eqnarray}
Summing (\ref{q1}), (\ref{q2}), (\ref{q3}), and (\ref{q4}), and using the following explicit formulas for the characteristic polynomial coefficients from \cite{det}
\begin{eqnarray}
&&b_{(1)}:=B+\widehat{\widetilde{B}}+\widehat{B}^\va+\widetilde{B}^\va\in\cl^0_{p,q},\nonumber\\
&&b_{(2)}:=-(B\widehat{\widetilde{B}}+B\widehat{B}^\va+B\widetilde{B}^\va+\widehat{\widetilde{B}}\widehat{B}^\va+ \widehat{\widetilde{B}}\widetilde{B}^\va+(\widehat{B}\widetilde{B})^\va)\in\cl^0_{p,q},\nonumber\\
&&b_{(3)}:=B \widehat{\widetilde{B}} \widehat{B}^\va+ B \widehat{\widetilde{B}} \widetilde{B}^\va+ B(\widehat{B} \widetilde{B})^\va+ \widehat{\widetilde{B}} (\widehat{B} \widetilde{B})^\va \in\cl^0_{p,q},\nonumber\\
&&b_{(4)}:=-\Det(B)=-B\widehat{\widetilde{B}}(\widehat{B} \widetilde{B})^{\va}\in\cl^0_{p,q},\nonumber
\end{eqnarray}
we get
\begin{eqnarray}
&&(A^4-A^3(B+\widehat{\widetilde{B}}+\widehat{B}^\va+\widetilde{B}^\va)\nonumber\\
&&+A^2(B\widehat{\widetilde{B}}+B\widehat{B}^\va+B\widetilde{B}^\va+\widehat{\widetilde{B}}\widehat{B}^\va+ \widehat{\widetilde{B}}\widetilde{B}^\va+(\widehat{B}\widetilde{B})^\va)\nonumber\\
&&-A(B \widehat{\widetilde{B}} \widehat{B}^\va+ B \widehat{\widetilde{B}} \widetilde{B}^\va+ B(\widehat{B} \widetilde{B})^\va+ \widehat{\widetilde{B}} (\widehat{B} \widetilde{B})^\va)+B\widehat{\widetilde{B}}(\widehat{B} \widetilde{B})^{\va})X\nonumber\\
&&=A^3 C-A^2C(\widehat{\widetilde{B}}+\widehat{B}^\va+\widetilde{B}^\va)\nonumber\\
&&+AC(\widehat{\widetilde{B}} \widehat{B}^\va+ \widehat{\widetilde{B}} \widetilde{B}^\va+ (\widehat{B} \widetilde{B})^\va)-C\widehat{\widetilde{B}}(\widehat{B} \widetilde{B})^{\va}.\nonumber
\end{eqnarray}
Denoting (\ref{D}) and (\ref{F}), and using the formula for the inverse in $\cl_{p,q}$ with $n=p+q=4$
$$
\Adj(D)=\widehat{\widetilde{D}}(\widehat{D} \widetilde{D})^{\va},\qquad \Det(D)=D\widehat{\widetilde{D}}(\widehat{D} \widetilde{D})^{\va},\qquad D^{-1}=\frac{\Adj(D)}{\Det(D)},\label{inv4}
$$
we obtain (\ref{X}).
\end{proof}

Let us present other formulas for the characteristic polynomial coefficients $b_{(1)}$, $b_{(2)}$, $b_{(3)}$, $b_{(4)}$ in the case $n=4$. We use the same expressions in the case $n=5$ (see Theorem \ref{thSylv5}). We have
\begin{eqnarray*}
B_{(1)}&:=&B,\\
b_{(1)}&=&4 \la B_{(1)} \ra_0=B+\widetilde{B}+\widehat{B}^\va+\widetilde{\widehat{B}}^\va,\\ B_{(2)}&:=&B(B-b_{(1)})=-B(\widetilde{B}+\widehat{B}^\va+\widetilde{\widehat{B}}^\va),\\
b_{(2)}&=&2 \la B_{(2)} \ra_0=-2 \la B(\widetilde{B}+\widehat{B}^\va+\widetilde{\widehat{B}}^\va) \ra_0\\
&=&-\frac{1}{2}(B\widetilde{B}+B\widehat{B}^\va+B\widetilde{\widehat{B}}^\va +B\widetilde{B}+ \widetilde{\widehat{B}}^\va \widetilde{B} +\widehat{B}^\va \widetilde{B}+(\widehat{B}\widetilde{\widehat{B}})^\va \\
&&+(\widehat{B} B^\va)^\va+(\widehat{B}\widetilde{B}^\va)^\va +(\widehat{B} \widetilde{\widehat{B}})^\va+(\widetilde{B}^\va \widetilde{\widehat{B}})^\va+(B^\va \widetilde{\widehat{B}})^\va)\\
&=&-(B\widetilde{B}+B\widehat{B}^\va+B\widetilde{\widehat{B}}^\va+\widetilde{B}\widehat{B}^\va+ \widetilde{B}\widetilde{\widehat{B}}^\va+(\widehat{B}\widetilde{\widehat{B}})^\va),\\
B_{(3)}&:=&B(B_{(2)}-b_{(2)})=B(\widetilde{B}\widehat{B}^\va+ \widetilde{B}\widetilde{\widehat{B}}^\va+(\widehat{B}\widetilde{\widehat{B}})^\va),\\
b_{(3)}&=&\frac{4}{3}\la B_{(3)} \ra_0=\frac{1}{3}(B \widetilde{B} \widehat{B}^\va+ B \widetilde{B}\widetilde{\widehat{B}}^\va+B(\widehat{B} \widetilde{\widehat{B}})^\va+\widetilde{\widehat{B}}^\va B \widetilde{B}\\
&&+\widehat{B}^\va B \widetilde{B}+(\widehat{B} \widetilde{\widehat{B}})^\va \widetilde{B}+(\widehat{B} \widetilde{\widehat{B}} B^\va)^\va+(\widehat{B}\widetilde{\widehat{B}}\widetilde{B}^\va)^\va\\
&&+ (\widehat{B} (B \widetilde{B})^\va)^\va+(\widetilde{B}^\va \widehat{B}\widetilde{\widehat{B}})^\va+ (B^\va \widehat{B} \widetilde{\widehat{B}})^\va+((B \widetilde{B})^\va \widetilde{\widehat{B}})^\va)\\
&=&B \widetilde{B} \widehat{B}^\va+ B \widetilde{B} \widetilde{\widehat{B}}^\va+ B(\widehat{B} \widetilde{\widehat{B}})^\va+ \widetilde{B} (\widehat{B} \widetilde{\widehat{B}})^\va,\\
\Det(B)&=&-B_{(4)}:=B(b_{(3)}-B_{(3)})=B\widetilde{B}(\widehat{B} \widetilde{\widehat{B}})^{\va}=-b_{(4)},
\end{eqnarray*}
where we used two times computer calculations\footnote{Analytic proof is also possible using the methods from \cite{Abd}.} to simplify the expressions for $b_{(2)}$ and $b_{(3)}$, because of nontrivial properties of the operation $\va$.

Instead of (\ref{X}), we obtain another equivalent form of basis-free solution to the Sylvester equation in the case $n=4$.
If
\begin{eqnarray}
Q:=D\widetilde{D}(\widehat{D} \widetilde{\widehat{D}})^{\va}\neq 0,\label{w1}
\end{eqnarray}
then
\begin{eqnarray}
X=\frac{\widetilde{D}(\widehat{D} \widetilde{\widehat{D}})^{\va} F}{Q},\label{w2}
\end{eqnarray}
where
\begin{eqnarray}
D&:=&A^4-A^3(B+\widetilde{B}+\widehat{B}^\va+\widetilde{\widehat{B}}^\va)\label{w3}\\
&+&A^2(B\widetilde{B}+B\widehat{B}^\va+B\widetilde{\widehat{B}}^\va+\widetilde{B}\widehat{B}^\va+ \widetilde{B}\widetilde{\widehat{B}}^\va+(\widehat{B}\widetilde{\widehat{B}})^\va)\nonumber\\
&-&A(B \widetilde{B} \widehat{B}^\va+ B \widetilde{B} \widetilde{\widehat{B}}^\va+ B(\widehat{B} \widetilde{\widehat{B}})^\va+ \widetilde{B} (\widehat{B} \widetilde{\widehat{B}})^\va)+B\widetilde{B}(\widehat{B} \widetilde{\widehat{B}})^{\va},\nonumber\\
F&:=&A^3 C-A^2C(\widetilde{B}+\widehat{B}^\va+\widetilde{\widehat{B}}^\va)\label{w4}\\
&+&AC(\widetilde{B}\widehat{B}^\va+ \widetilde{B}\widetilde{\widehat{B}}^\va+(\widehat{B}\widetilde{\widehat{B}})^\va)-C\widetilde{B}(\widehat{B} \widetilde{\widehat{B}})^{\va}.\nonumber
\end{eqnarray}
We use the same expressions in the case $n=5$ (see the next section).

Note that the formulas (\ref{w1}), (\ref{w2}), (\ref{w3}), and (\ref{w4}) can be rewritten using the new operation
\begin{eqnarray}
B^\sharp:=(\widehat{B}\widetilde{\widehat{B}})^\va\label{sharp}
\end{eqnarray}
in the form
\begin{eqnarray*}
&&Q=D\widetilde{D}D^\sharp,\qquad
X=\frac{\widetilde{D}D^\sharp F}{Q},\\
&&D=A^4-A^3(B+\widetilde{B}+\widehat{B}^\va+\widetilde{\widehat{B}}^\va)+ A^2(B\widetilde{B}+B\widehat{B}^\va+B\widetilde{\widehat{B}}^\va+\widetilde{B}\widehat{B}^\va\\
&&+ \widetilde{B}\widetilde{\widehat{B}}^\va+B^\sharp)
-A(B \widetilde{B} \widehat{B}^\va+ B \widetilde{B} \widetilde{\widehat{B}}^\va+ BB^\sharp+ \widetilde{B} B^\sharp)+B\widetilde{B}B^\sharp,\\
&&F=A^3 C-A^2C(\widetilde{B}+\widehat{B}^\va+\widetilde{\widehat{B}}^\va)
+AC(\widetilde{B}\widehat{B}^\va+ \widetilde{B}\widetilde{\widehat{B}}^\va+B^\sharp)-C\widetilde{B}B^\sharp.
\end{eqnarray*}

\begin{ex} Let us consider the Sylvester equation (\ref{Sylv4}) in $\cl_{1,3}$ with
\begin{eqnarray*}
A&=&3e- 5e_1+ 2 e_2 - 2 e_3 - 4 e_4 + e_{12} +
 3 e_{13} + 5 e_{14} + 2 e_{23} + 2 e_{24} -
 5 e_{34}\\
&& + 2 e_{123} - 4 e_{124} + e_{134} +
 4 e_{234} + 2 e_{1234},\\
 B&=&2 e + 5 e_1 - e_2 - 2 e_3 - e_4 + e_{12} +
 2 e_{13} + 5 e_{14} - 5 e_{23} + 2 e_{24} -
 3 e_{34}\\
 &&+ 4 e_{123} - 3 e_{124} + 4 e_{134} +
 3 e_{234} + e_{1234},\\
 C&=&4 e + e_1- 3 e_2 - 2 e_3 + 4 e_4 - e_{12} +
 5 e_{13} + e_{14}+ 3 e_{23} + e_{24}- 4 e_{34}\\
 && +2 e_{123} - 3 e_{124} - 2 e_{134} - 5 e_{234} +
 5 e_{1234}.
 \end{eqnarray*}
Using the formulas (\ref{w1}), (\ref{w2}), (\ref{w3}), (\ref{w4}), and computer calculations in Wolfram Mathematica, we get
\begin{eqnarray}
D&=&-3331 e + 16960 e_1 - 2736 e_2+ 5228 e_3 +
 11276 e_4 - 4372 e_{12}- 6740 e_{13}\nonumber\\
 &&- 17764 e_{14} -
 4208 e_{23} - 4520 e_{24} + 12072 e_{34} -
 8664 e_{123} + 8664 e_{124} \nonumber\\
 &&- 2128 e_{134} -
 14968 e_{234} - 5868 e_{1234},\nonumber\\
 Q&=&818014056354052817 e\neq 0,\nonumber\\
 F&=&-3654 e - 3114 e_1- 4238 e_2- 12909 e_3 - 629 e_4 -
 7164 e_{12} - 5583 e_{13}\nonumber\\
 && - 9442 e_{14} - 14155 e_{23} -
 1197 e_{24} + 3316 e_{34} - 9352 e_{123} -
 2768 e_{124},\nonumber\\
 && - 2570 e_{134} + 6614 e_{234} -
 6485 e_{1234},\nonumber\\
 X&=&\frac{1}{Q}(-119559672248263574 e - 243271127103539030 e_1\label{w5}\\
 &&-  45110505690078854 e_2 + 102025493907271711 e_3 \nonumber\\
 &&-  237419769499231033 e_4 - 230234896037415164 e_{12}\nonumber\\
 && -  631822395022405163 e_{13} + 354830063944470830 e_{14}\nonumber\\
 && -  248262081322178503 e_{23} + 381628355781437695 e_{24}\nonumber\\
 && +  242277961566965860 e_{34} + 175205777213912492 e_{123}\nonumber\\
 && +  85615763017907532 e_{124} - 78264759152759606 e_{134}\nonumber\\
 && + 12173556035563862 e_{234} + 268142275333252559 e_{1234}).\nonumber
\end{eqnarray}
Substituting (\ref{w5}) into (\ref{Sylv4}), we get equality.
\end{ex}

\section{The Case $n=5$}

\begin{thm}\label{thSylv5} Let us consider the Sylvester equation in $\cl_{p,q}$, $p+q=5$,
\begin{eqnarray}
AX-XB=C\label{Sylv5}
\end{eqnarray}
for known $A, B, C\in\cl_{p,q}$ and unknown $X\in\cl_{p,q}$.

If
\begin{eqnarray}
Q:=D\widetilde{D}(\widehat{D} \widehat{\widetilde{D}})^{\va} (D\widetilde{D}(\widehat{D} \widehat{\widetilde{D}})^{\va})^\va\neq 0,\label{Q5}
\end{eqnarray}
then
\begin{eqnarray}
X=\frac{\widetilde{D}(\widehat{D} \widehat{\widetilde{D}})^{\va} (D\widetilde{D}(\widehat{D} \widehat{\widetilde{D}})^{\va})^\va F}{Q},\label{sol5}
\end{eqnarray}
where
\begin{eqnarray}
D&:=&A^4-A^3(B+\widetilde{B}+\widehat{B}^\va+\widetilde{\widehat{B}}^\va)\label{D5}\\
&+&A^2(B\widetilde{B}+B\widehat{B}^\va+B\widetilde{\widehat{B}}^\va+\widetilde{B}\widehat{B}^\va+ \widetilde{B}\widetilde{\widehat{B}}^\va+(\widehat{B}\widetilde{\widehat{B}})^\va)\nonumber\\
&-&A(B \widetilde{B} \widehat{B}^\va+ B \widetilde{B} \widetilde{\widehat{B}}^\va+ B(\widehat{B} \widetilde{\widehat{B}})^\va+ \widetilde{B} (\widehat{B} \widetilde{\widehat{B}})^\va)+B\widetilde{B}(\widehat{B} \widetilde{\widehat{B}})^{\va},\nonumber\\
F&:=&A^3 C-A^2C(\widetilde{B}+\widehat{B}^\va+\widetilde{\widehat{B}}^\va)\label{F5}\\
&+&AC(\widetilde{B}\widehat{B}^\va+ \widetilde{B}\widetilde{\widehat{B}}^\va+(\widehat{B}\widetilde{\widehat{B}})^\va)-C\widetilde{B}(\widehat{B} \widetilde{\widehat{B}})^{\va}.\nonumber
\end{eqnarray}
\end{thm}
Note that the formulas (\ref{Q5}), (\ref{sol5}), (\ref{D5}), and (\ref{F5}) can be rewritten using the operation (\ref{sharp}) in the form
\begin{eqnarray*}
&&Q=D\widetilde{D}D^\sharp (D\widetilde{D}D^\sharp)^\va,\qquad
X=\frac{\widetilde{D}D^\sharp (D\widetilde{D}D^\sharp)^\va F}{Q},\\
&&D=A^4-A^3(B+\widetilde{B}+\widehat{B}^\va+\widetilde{\widehat{B}}^\va)+ A^2(B\widetilde{B}+B\widehat{B}^\va+B\widetilde{\widehat{B}}^\va+\widetilde{B}\widehat{B}^\va\\
&&+ \widetilde{B}\widetilde{\widehat{B}}^\va+B^\sharp)
-A(B \widetilde{B} \widehat{B}^\va+ B \widetilde{B} \widetilde{\widehat{B}}^\va+ BB^\sharp+ \widetilde{B} B^\sharp)+B\widetilde{B}B^\sharp,\\
&&F=A^3 C-A^2C(\widetilde{B}+\widehat{B}^\va+\widetilde{\widehat{B}}^\va)
+AC(\widetilde{B}\widehat{B}^\va+ \widetilde{B}\widetilde{\widehat{B}}^\va+B^\sharp)-C\widetilde{B}B^\sharp.
\end{eqnarray*}

\begin{proof} In the case $\cl_{p,q}$, $p+q=5$, we have 8 characteristic polynomial coefficients\footnote{Explicit formulas for the coefficients $b_{(1)}$, \ldots, $b_{(8)}$ are presented in \cite{Abd}.} $b_{(1)}$, \ldots, $b_{(8)}$ for an arbitrary element $B\in\cl_{p,q}$. Instead of them, let us consider the following 4 expressions (which are scalars in the case $n=4$)
\begin{eqnarray}
&&b^\prime_{(1)}=B+\widetilde{B}+\widehat{B}^\va+\widetilde{\widehat{B}}^\va,\nonumber\\
&&b^\prime_{(2)}=-(B\widetilde{B}+B\widehat{B}^\va+B\widetilde{\widehat{B}}^\va+\widetilde{B}\widehat{B}^\va+ \widetilde{B}\widetilde{\widehat{B}}^\va+(\widehat{B}\widetilde{\widehat{B}})^\va),\label{RR}\\
&&b^\prime_{(3)}=B \widetilde{B} \widehat{B}^\va+ B \widetilde{B} \widetilde{\widehat{B}}^\va+ B(\widehat{B} \widetilde{\widehat{B}})^\va+ \widetilde{B} (\widehat{B} \widetilde{\widehat{B}})^\va,\nonumber\\
&&b^\prime_{(4)}=-B\widetilde{B}(\widehat{B} \widetilde{\widehat{B}})^{\va}.\nonumber
\end{eqnarray}
We have\footnote{Note that the same is not true for the expressions for characteristic polynomial coefficients from Theorem \ref{thSylv4}. For example, $B\widehat{\widetilde{B}}(\widehat{B} \widetilde{B})^{\va}\in\cl^0_{p,q}\oplus\cl^1_{p,q}\oplus\cl^4_{p,q}\neq \cen(\cl_{p,q})$ in the case $n=5$, see the details in \cite{det}.} $b^\prime_{(1)}, b^\prime_{(2)}, b^\prime_{(3)}, b^\prime_{(4)}\in\cen(\cl_{p,q})=\cl^0_{p,q}\oplus\cl^5_{p,q}$. We can easily verify that
$$B+\widetilde{B}+\widehat{B}^\va+\widetilde{\widehat{B}}^\va=4(\la B \ra_0+ \la B \ra_5)\in\cl^0_{p,q}\oplus\cl^5_{p,q}$$
using definitions of the operations (\ref{hat}), (\ref{tilde}), and (\ref{tri}). We have  $B\widetilde{B}(\widehat{B} \widetilde{\widehat{B}})^{\va}\in\cl^0_{p,q}\oplus\cl^5_{p,q}$ (see \cite{det}). We verified $b^\prime_{(2)}, b^\prime_{(3)}\in\cl^0_{p,q}\oplus\cl^5_{p,q}$ using computer calculations\footnote{Analytic proof is also possible using the methods from \cite{Abd}.} .

Multiplying both sides of (\ref{Sylv5}) on the right by $-\widetilde{B}(\widehat{B} \widetilde{\widehat{B}})^{\va}$, we get
\begin{eqnarray}
-AX\widetilde{B}(\widehat{B} \widetilde{\widehat{B}})^{\va}+X B\widetilde{B}(\widehat{B} \widetilde{\widehat{B}})^{\va}=-C\widetilde{B}(\widehat{B} \widetilde{\widehat{B}})^{\va}.\label{q15}
\end{eqnarray}
Multiplying both sides of (\ref{Sylv5}) on the right by $\widetilde{B}\widehat{B}^\va+ \widetilde{B} \widetilde{\widehat{B}}^\va+ (\widehat{B} \widetilde{\widehat{B}})^\va$  and on the left by $A$, we get
\begin{eqnarray}
&&A^2 X (\widetilde{B}\widehat{B}^\va+ \widetilde{B} \widetilde{\widehat{B}}^\va+ (\widehat{B} \widetilde{\widehat{B}})^\va) -AXB(\widetilde{B}\widehat{B}^\va+ \widetilde{B} \widetilde{\widehat{B}}^\va+ (\widehat{B} \widetilde{\widehat{B}})^\va)\nonumber\\
&&=AC(\widetilde{B}\widehat{B}^\va+ \widetilde{B} \widetilde{\widehat{B}}^\va+ (\widehat{B} \widetilde{\widehat{B}})^\va).\label{q25}
\end{eqnarray}
Multiplying both sides of (\ref{Sylv5}) on the right by $-(\widetilde{B}+\widehat{B}^\va+\widetilde{\widehat{B}}^\va)$ and on the left by $A^2$, we get
\begin{eqnarray}
&&-A^3 X(\widetilde{B}+\widehat{B}^\va+\widetilde{\widehat{B}}^\va)+ A^2 XB(\widetilde{B}+\widehat{B}^\va+\widetilde{\widehat{B}}^\va)\nonumber\\
&&= -A^2C(\widetilde{B}+\widehat{B}^\va+\widetilde{\widehat{B}}^\va).\label{q35}
\end{eqnarray}
Multiplying both sides of (\ref{Sylv5}) on the left by $A^3$, we get
\begin{eqnarray}
A^4X-A^3 XB=A^3C.\label{q45}
\end{eqnarray}
Summing (\ref{q15}), (\ref{q25}), (\ref{q35}), and (\ref{q45}), we get
\begin{eqnarray}
&&(A^4-A^3(B+\widetilde{B}+\widehat{B}^\va+\widetilde{\widehat{B}}^\va)\nonumber\\
&&+A^2(B\widetilde{B}+B\widehat{B}^\va+B\widetilde{\widehat{B}}^\va+\widetilde{B}\widehat{B}^\va+ \widetilde{B}\widetilde{\widehat{B}}^\va+(\widehat{B}\widetilde{\widehat{B}})^\va)\nonumber\\
&&-A(B \widetilde{B} \widehat{B}^\va+ B \widetilde{B} \widetilde{\widehat{B}}^\va+ B(\widehat{B} \widetilde{\widehat{B}})^\va+ \widetilde{B} (\widehat{B} \widetilde{\widehat{B}})^\va)+B\widetilde{B}(\widehat{B} \widetilde{\widehat{B}})^{\va})X\nonumber\\
&&=A^3 C-A^2C(\widetilde{B}+\widehat{B}^\va+\widetilde{\widehat{B}}^\va)\nonumber\\
&&+AC(\widetilde{B}\widehat{B}^\va+ \widetilde{B}\widetilde{\widehat{B}}^\va+(\widehat{B}\widetilde{\widehat{B}})^\va)-C\widetilde{B}(\widehat{B} \widetilde{\widehat{B}})^{\va}.\nonumber
\end{eqnarray}
Denoting (\ref{D5}) and (\ref{F5}), and using the formula for the inverse in the case $n=5$ (see \cite{det}):
\begin{eqnarray*}
&&D^{-1}=\frac{\Adj(D)}{\Det(D)},\qquad \Det(D)=D\widetilde{D}(\widehat{D} \widehat{\widetilde{D}})^{\va} (D\widetilde{D}(\widehat{D} \widehat{\widetilde{D}})^{\va})^\va\in\cl^0_{p,q}\equiv \R,\\
&&\Adj(D)=\widetilde{D}(\widehat{D} \widehat{\widetilde{D}})^{\va} (D\widetilde{D}(\widehat{D} \widehat{\widetilde{D}})^{\va})^\va,
\end{eqnarray*}
we get (\ref{sol5}).
\end{proof}

\begin{ex} Let us consider the Sylvester equation (\ref{Sylv5}) in $\cl_{4,1}$ with
\begin{eqnarray*}
A&=&- e + e_1 - 3 e_2 + 3 e_3 + 2 e_4 - e_5 +
 3 e_{12} + 2 e_{13} -2 e_{14} -3 e_{15} -
 e_{23}\\
 && - 3 e_{24} - e_{25} - e_{34}  -3 e_{35} -
 3 e_{45} - e_{123} - 3 e_{124} + e_{125} -
 e_{134}  -3 e_{135}\\
 && + e_{145}+ 2 e_{234} +
 2 e_{235}  -2 e_{245} + 3 e_{345} +
 3 e_{1234} - 2 e_{1235} + 2 e_{1245}\\
 && - e_{1345} -2 e_{2345} - 2 e_{12345},\\
 B&=&-2 e - e_1 - 3 e_2 - 2 e_3 -  e_4 + e_5 -
 2 e_{12} + 2 e_{13} - e_{14} - 2 e_{15} + 3 e_{23}\\
 && +e_{24} - 2 e_{25}- 3 e_{34} + 2 e_{35} - 3 e_{45} -
 e_{123} + e_{124} + 2 e_{125} - 2 e_{134} +
 3 e_{135}\\
 && + 3 e_{145} - 3 e_{234} - e_{235} -
 3 e_{245} - e_{345} + e_{1234} + e_{1235} +
 e_{1245}\\
 && + 3 e_{1345} + 2 e_{2345} -
 3 e_{12345},\\
 C&=&3 e - 3 e_1 + 2 e_2 + e_3 + 3 e_4 - 2 e_5 -
 3 e_{12} - 2 e_{13} + e_{14} - e_{15} + 2 e_{23} \\
 &&+
 2 e_{24} + 2 e_{25}- 2 e_{34} - 3 e_{35} + e_{45} +
 e_{123} - 3 e_{124} + e_{125} - e_{134} -
 2 e_{135}\\
 && - 2 e_{145} - e_{234} - 2 e_{235} -
 3 e_{245} - 2 e_{345} - e_{1234} -
 3 e_{1235} + e_{1245}\\
 && - 2 e_{1345} -
 e_{2345} - e_{12345}.
 \end{eqnarray*}
Using the formulas (\ref{Q5}), (\ref{sol5}), (\ref{D5}), (\ref{F5}), and computer calculations in Wolfram Mathematica, we get
\begin{eqnarray}
D&=&-28 e - 2784 e_1 + 4088 e_2 - 1584 e_3 + 1432 e_4 -
 2528 e_5- 1688 e_{12}\nonumber\\
 &&- 2496 e_{13} - 3624 e_{14} -
 1392 e_{15} + 2904 e_{23} - 3392 e_{24} - 648 e_{25}\nonumber \\
 &&+
 1664 e_{34} + 3104 e_{35} + 3088 e_{45} +
 1568 e_{123} - 3056 e_{124} - 4272 e_{125}\nonumber\\
 && +
 2888 e_{134} + 4280 e_{135} + 296 e_{145} +
 416 e_{234} - 936 e_{235} + 560 e_{245} \nonumber\\
 &&+
 3064 e_{345} - 208 e_{1234} + 4648 e_{1235} -
 1632 e_{1245} - 1528 e_{1345}\nonumber\\
 && - 1712 e_{2345} -
 1112 e_{12345},\nonumber\\
 Q&=&269517633593422176823514562560 e\neq 0,\nonumber\\
 F&=&-4792 e - 4250 e_1 + 2398 e_2+ 1168 e_3 - 8208 e_4 +
 3268 e_5 + 784 e_{12}\nonumber \\
 &&+ 4594 e_{13} + 2108 e_{14} -
 4948 e_{15} - 1454 e_{23}+ 606 e_{24} + 2350 e_{25}\nonumber \\
 &&+
 7786 e_{34} - 3102 e_{35} + 8970 e_{45} -
 10044 e_{123} - 4682 e_{124} + 5594 e_{125}\nonumber \\
 &&+
 3822 e_{134} - 1034 e_{135} + 9688 e_{145} +
 6272 e_{234} + 4448 e_{235} + 5580 e_{245}\nonumber\\
 && -
 222 e_{345} + 3816 e_{1234} + 2648 e_{1235} +
 7042 e_{1245} + 6824 e_{1345}\nonumber\\
 && + 6496 e_{2345} -
 5290 e_{12345},\nonumber
 \end{eqnarray}
 and
{\scriptsize
 \begin{eqnarray}
 X&=&\frac{1}{Q}(-254263734302655483397831852032 e +
 124333161192922434122282795008 e_1 \label{q5}\\
 &&+
 4254232860869616089214910464 e_2 +
 77590614000116777995555176448 e_3 \nonumber\\
 &&-
 274797689363873365890872967168 e_4 -
 251661656524140523469539442688 e_5\nonumber \\
 &&+
 172171077855495001058426880000 e_{12} +
 450974610831748898553901056000 e_{13} \nonumber\\
 &&+
 125362877685993488610803777536 e_{14} -
 105084562335627847197682171904 e_{15}\nonumber\\
 && -
 369396757288051245240066080768 e_{23} +
 24152800954960269837389037568 e_{24}\nonumber\\
 && +
 404037705534519977662329880576 e_{25} +
 154018345680240445994135486464 e_{34}\nonumber \\
&& -
 75749500716946413019089633280 e_{35} -
 65907834985445771184840605696 e_{45}\nonumber \\
&& -
 350045165524614842201912639488 e_{123} -
 103045468655912395414190981120 e_{124}\nonumber \\
&& +
 590953599976818750339972169728 e_{125} +
 73707775121065150382774059008 e_{134}\nonumber\\
 && -
 231126475215498004282582532096 e_{135} +
 271845290646123969359373860864 e_{145}\nonumber\\
 && -
 106624544616437709970075025408 e_{234}+
 148392611724890260094364352512 e_{235}\nonumber\\
 &&+
 82665842802946364223515852800 e_{245} +
 37807107639770035301829672960 e_{345}\nonumber \\
 &&+
 220607526593898040150924263424 e_{1234} -
 122610796966295721009684545536 e_{1235}\nonumber \\
 &&+
 200466518963156538449965973504 e_{1245} +
 273731675174703848612170170368 e_{1345}\nonumber \\
 &&+
 249675343428108665838275067904 e_{2345} -
 345584811030745943796431486976 e_{12345}).\nonumber
\end{eqnarray}}
Substituting (\ref{q5}) into (\ref{Sylv5}), we get equality.
\end{ex}

\section{The Case of Arbitrary $n$}

Let us consider the general case of the real Clifford algebra $\cl_{p,q}$ with arbitrary $n=p+q$. We use the following concepts of characteristic polynomial $\varphi_B(\lambda)$, determinant $\Det(B)$, adjugate $\Adj(B)$, and inverse $B^{-1}$ in $\cl_{p,q}$ (see the details in \cite{det})\footnote{Here and below we denote the integer part of the number $\frac{n+1}{2}$ by $[\frac{n+1}{2}]$.}:
\begin{eqnarray*}
&&\varphi_B(\lambda):=\Det(\lambda e-B)=\lambda^N-b_{(1)}\lambda^{N-1}-\cdots-b_{(N)}\in\cl^0_{p,q},\\
&&B_{(1)}:=B,\quad B_{(k+1)}:=B(B_{(k)}-b_{(k)}),\qquad N:=2^{[\frac{n+1}{2}]},\\
&&b_{(k)}=\frac{N}{k} \la B_{(k)}\ra_0\in\cl^0_{p,q}\equiv \R,\qquad k=1, \ldots, N,\\
&&\Det(B)=-B_{(N)}=-b_{(N)}=B(b_{(N-1)}-B_{(N-1)})\in\cl^0_{p,q}\equiv \R,\\
&&\Adj(B)=b_{(N-1)}-B_{{(N-1)}},\qquad B^{-1}=\frac{\Adj(B)}{\Det(B)}.
\end{eqnarray*}

In the following theorem, we present recursive formulas for the basis-free solution to the Sylvester equation in the case of arbitrary $n=p+q$.

\begin{thm}\label{thgen} Let us consider the Sylvester equation in $\cl_{p,q}$, $p+q=n$,
\begin{eqnarray}
AX-XB=C\label{Sylvgen}
\end{eqnarray}
for known $A, B, C\in\cl_{p,q}$ and unknown $X\in\cl_{p,q}$.

Let us denote $N:=2^{[\frac{n+1}{2}]}$. If
\begin{eqnarray}
Q:=d_{(N)}\neq 0,\label{Qn}
\end{eqnarray}
then
\begin{eqnarray}
X=\frac{(D_{(N-1)}-d_{(N-1)})F}{Q},\label{X3}
\end{eqnarray}
where
\begin{eqnarray}
&&D:=-\sum_{j=0}^N A^{N-j} b_{(j)},\label{Dgen}\\
&&F:=\sum_{j=1}^{N} A^{N-j} C (B_{(j-1)}-b_{(j-1)}),\label{Fgen}
\end{eqnarray}
and the following expressions are defined recursively
\footnote{Note that using the recursive formulas $B_{(k+1)}=B (B_{(k)}-b_{(k)})$, the expression (\ref{Fgen}) can be reduced to the form $\sum_{i, j} b_{ij} A^i C B^j$ with some scalars $b_{ij}\in\R$.}
:
\begin{eqnarray*}
&&b_{(k)}=\frac{N}{k}\la B_{(k)}\ra_0,\quad B_{(k+1)}=B (B_{(k)}-b_{(k)}),\quad B_{(1)}=B,\\
&&d_{(k)}=\frac{N}{k}\la D_{(k)}\ra_0,\quad D_{(k+1)}=D (D_{(k)}-d_{(k)}),\quad D_{(1)}=D,\\
&&B_{(0)}=D_{(0)}:=0,\qquad b_{(0)}=d_{(0)}:=-1,\qquad k=1,\ldots, N.
\end{eqnarray*}
\end{thm}

Note that $D$ (\ref{Dgen}) is the characteristic polynomial of the element $B$ with the substitution of $A$.

\begin{proof} Multiplying both sides of (\ref{Sylvgen}) on the right by $B_{(N-1)}-b_{(N-1)}$, on the right by $B_{(N-2)}-b_{(N-2)}$ and on the left by $A$, on the right by $B_{(N-3)}-b_{(N-3)}$ and on the left by $A^2$, \ldots, on the right by $B_{(2)}-b_{(2)}$ and on the left by $A^{N-3}$, on the right by $B-b_{(1)}$ and on the left by $A^{N-2}$, on the left by $A^{N-1}$, we get
\begin{eqnarray*}
&&AX(B_{(N-1)}-b_{(N-1)})-XB(B_{(N-1)}-b_{(N-1)})\\
&&=C(B_{(N-1)}-b_{(N-1)}),\\
&&A^2 X (B_{(N-2)}-b_{(N-2)})-AXB(B_{(N-2)}-b_{(N-2)})\\
&&=AC(B_{(N-2)}-b_{(N-2)}),\\
&&A^3 X(B_{(N-3)}-b_{(N-3)})-A^2XB(B_{(N-3)}-b_{(N-3)})\\
&&=A^2 C (B_{(N-3)}-b_{(N-3)}),\\
&&\cdots\\
&& A^{N-2}X(B_{(2)}-b_{(2)})-A^{N-3}XB (B_{(2)}-b_{(2)})=A^{N-3}C (B_{(2)}-b_{(2)}),\\
&&A^{N-1} X (B-b_{(1)})-A^{N-2}XB(B-b_{(1)})=A^{N-2}C(B-b_{(1)}),\\
&&A^{N} X- A^{N-1}XB=A^{N-1}C.
\end{eqnarray*}
Summing these equations and using
\begin{eqnarray*}
&&B_{(k+1)}=B (B_{(k)}-b_{(k)}),\qquad k=1, \ldots, N,\\
&&B_{(N)}=B(B_{(N-1)}-b_{(N-1)})=b_{(N)}=-\Det(B)\in\cl^0_{p,q},
\end{eqnarray*}
we get
\begin{eqnarray*}
&&(A^N-b_{(1)}A^{N-1}X-b_{(2)}A^{N-2}X-\cdots-b_{(N-1)}AX-b_{(N)})X\\
&&=A^{N-1}C+A^{N-2}C(B-b_{(1)})+\cdots+C(B_{(N-1)}-b_{(N-1)}).
\end{eqnarray*}
Denoting (\ref{Dgen}), (\ref{Fgen}), and using
$$D^{-1}=\frac{\Adj(D)}{\Det(D)},\quad \Adj(D)=d_{(N-1)}-D_{{(N-1)}},\quad \Det(D)=-d_{(N)},$$
we get (\ref{X3}).
\end{proof}

Note that in the case of odd $n$, the formulas (\ref{Dgen}) and (\ref{Fgen}) can be simplified. We can use instead of $N$ characteristic polynomial coefficients some other $\frac{N}{2}$ expressions. We call them  generalized characteristic polynomial coefficients. For example, in the case $n=5$ (see Theorem \ref{thSylv5}), we use the $4$ expressions $b^\prime_{(k)}$, $k=1, 2, 3, 4$, (\ref{RR}), which are in the center of $\cl_{p,q}$, instead of the $8$ ordinary characteristic polynomial coefficients $b_{(k)}$, $k=1, \ldots, 8$, which are in $\cl^0_{p,q}$.

The ordinary characteristic polynomial coefficients of Clifford algebra element corresponds to the characteristic polynomial coefficients of the corresponding matrix representation of dimension $N$ (see the details in \cite{det}). The generalized characteristic polynomial coefficients of Clifford algebra element corresponds to the characteristic polynomial coefficients of the corresponding matrix of dimension $\frac{N}{2}$ with entries in $\C$ or $\R\oplus\R$. In more details, the center of $\cl_{p,q}$ with odd $n=p+q$ is $\cen(\cl_{p,q})=\cl^0_{p,q}\oplus\cl^n_{p,q}$, which is isomorphic to $\C$ in the case $e_{1\ldots n}^2=-e$ (i.e. $p-q=2, 3\mod 4$) and to $\R\oplus\R$ in the case $e_{1\ldots n}^2=e$ (i.e. $p-q=0,1\mod 4$). The Clifford algebra $\cl_{p,q}$ with odd $n=p+q$ can be represented in the form (the same idea is used in \cite{hitzer2})
$$
\cl_{p,q}=\cl^{(0)}_{p,q}\oplus \cl^{(1)}_{p,q}=\cl^{(0)}_{p,q}\oplus e_{1\ldots n}\cl^{(0)}_{p,q},
$$
where $$\cl^{(0)}_{p,q}=\bigoplus_{k=0 \mod 2} \cl^k_{p,q},\qquad \cl^{(1)}_{p,q}=\bigoplus_{k=1\mod 2}\cl^k_{p,q}$$
are the even subalgebra and the odd subspace of $\cl_{p,q}$. Thus any element $B\in\cl_{p,q}$ can be written as an element of the even subalgebra $\cl^{(0)}_{p,q}$ with complex (in the cases $p-q=2, 3\mod 4$) or hyperbolic (in the cases $p-q=0, 1\mod 4$) coefficients. Also we use the well-known isomorphisms (see, for example, \cite{Lounesto, Bulg}) $$\cl^{(0)}_{p, q}\cong\cl_{p, q-1},\qquad q\geq 1;\qquad \cl^{(0)}_{p, q}\cong\cl_{q,p-1},\qquad p\geq 1.$$

We obtain the following simplification of the statement of Theorem \ref{thgen} in the case of odd $n=p+q$ (with $\frac{N}{2}$ steps in the corresponding recursive formulas for $D$ and $F$ instead of $N$ steps for these expressions as in the previous theorem).

Let us consider the Sylvester equation in $\cl_{p,q}$ with odd $p+q=n$,
\begin{eqnarray}
AX-XB=C
\end{eqnarray}
for known $A, B, C\in\cl_{p,q}$ and unknown $X\in\cl_{p,q}$. Let us denote\footnote{In the case of odd $n$, the integer part of the number $\frac{n+1}{2}$ is equal to $[\frac{n+1}{2}]=\frac{n+1}{2}\in{\mathbb Z}$.} $N:=2^{\frac{n+1}{2}}$. If
\begin{eqnarray}
Q:=d_{(N)}\neq 0,\label{Qnodd}
\end{eqnarray}
then
$$X=\frac{(D_{(N-1)}-d_{(N-1)})F}{d_{(N)}},$$
where
\begin{eqnarray}
&&D:=\varphi^\prime_B(A)=-\sum_{j=0}^{\frac{N}{2}} A^{\frac{N}{2}-j} b^\prime_{(j)},\label{Dgen2}\\
&&F:=\sum_{j=1}^{\frac{N}{2}} A^{\frac{N}{2}-j} C (B^\prime_{(j-1)}-b^\prime_{(j-1)}),\label{Fgen2}
\end{eqnarray}
and the following expressions are defined recursively:
\begin{eqnarray*}
&&b^\prime_{(k)}=\frac{N}{k}\la B^\prime_{(k)}\ra_{\cen},\quad B^\prime_{(k+1)}=B (B^\prime_{(k)}-b^\prime_{(k)}),\quad  k=1,\ldots, \frac{N}{2},\\
&&B^\prime_{(1)}=B,\qquad B^\prime_{(0)}:=0,\qquad b^\prime_{(0)}:=-1,\\
&&d_{(m)}=\frac{N}{m}\la D_{(m)}\ra_0,\quad D_{(m+1)}=D (D_{(m)}-d_{(m)}),\quad m=1, \ldots, N,\\
&&D_{(1)}=D,\quad D_{(0)}:=0,\quad d_{(0)}:=-1.
\end{eqnarray*}

Note that expressions for generalized characteristic polynomial coefficients $b^\prime_{(k)}$, $k=1, \ldots, \frac{N}{2}$ coincide with the expressions for the characteristic polynomial coefficients for the previous even $n-1$ (see the example for the case $n=5$ in Theorem \ref{thSylv5}).

\begin{ex}
Let us present the explicit expressions for $\la B \ra_{\cen}$ in the case of small dimensions $n\leq 15$ using the operations of conjugation $\quad\widetilde{}$,\quad $\widehat{}$,\, ${}^\va$,\, ${}^\square$\,:
\begin{eqnarray*}
&&\frac{1}{2}(B+\widetilde{\widehat{B}})=\la B \ra_{\cen}=\left\lbrace
\begin{array}{ll}
\la B \ra_0, & \mbox{if $n=2$,}\\
\la B \ra_0+\la B\ra_3, & \mbox{if $n=3$;}
\end{array}
\right.\\
&&\frac{1}{4}(B+\widetilde{B}+\widehat{B}^\va+\widetilde{\widehat{B}}^\va)=\la B \ra_{\cen}=\left\lbrace
\begin{array}{ll}
\la B \ra_0, & \mbox{if $n=4$,}\\
\la B \ra_0+\la B\ra_5, & \mbox{if $n=5$;}
\end{array}
\right.\\
&&\frac{1}{4}(B+\widehat{\widetilde{B}}+\widehat{B}^\va+\widetilde{B}^\va)=\la B \ra_{\cen}=\left\lbrace
\begin{array}{ll}
\la B \ra_0, & \mbox{if $n=6$,}\\
\la B \ra_0+\la B\ra_7, & \mbox{if $n=7$;}
\end{array}
\right.\\
&&\frac{1}{8}(B+\widehat{B}^{\square}+\widetilde{B}+\widehat{\widetilde{B}}^{\square}+B^{\va}+\widehat{B}^{\va\square}+\widetilde{B}^\va +\widehat{\widetilde{B}}^{\va\square})\\
&&=\la B \ra_{\cen}=\left\lbrace
\begin{array}{ll}
\la B \ra_0, & \mbox{if $n=8$,}\\
\la B \ra_0+\la B\ra_9, & \mbox{if $n=9$;}
\end{array}
\right.\\
&&\frac{1}{8}(B+\widehat{B}^{\square}+\widetilde{B}^\square+\widehat{\widetilde{B}}+B^{\va}+\widehat{B}^{\va\square}+\widetilde{B}^{\va\square} +\widehat{\widetilde{B}}^{\va\square})\\
&&=\la B \ra_{\cen}=\left\lbrace
\begin{array}{ll}
\la B \ra_0, & \mbox{if $n=10$,}\\
\la B \ra_0+\la B\ra_{11}, & \mbox{if $n=11$;}
\end{array}
\right.\\
&&\frac{1}{8}(B+\widehat{B}^{\square}+\widetilde{B}+\widehat{\widetilde{B}}^\square+B^{\va\square}+\widehat{B}^{\va}+\widetilde{B}^{\va\square} +\widehat{\widetilde{B}}^{\va})\\
&&=\la B \ra_{\cen}=\left\lbrace
\begin{array}{ll}
\la B \ra_0, & \mbox{if $n=12$,}\\
\la B \ra_0+\la B\ra_{13}, & \mbox{if $n=13$;}
\end{array}
\right.\\
&&\frac{1}{8}(B+\widehat{B}^{\square}+\widetilde{B}^\square+\widehat{\widetilde{B}}+B^{\va\square}+\widehat{B}^{\va}+\widetilde{B}^{\va} +\widehat{\widetilde{B}}^{\va\square})\\
&&=\la B \ra_{\cen}=\left\lbrace
\begin{array}{ll}
\la B \ra_0, & \mbox{if $n=14$,}\\
\la B \ra_0+\la B\ra_{15}, & \mbox{if $n=15$,}
\end{array}
\right.\\
\end{eqnarray*}
where we use the additional operation of conjugation $\, {}^\square\,$ (compare with (\ref{hat}), (\ref{tilde}), and (\ref{tri}), see also \cite{det})
\begin{eqnarray*}
B^\square&=&\sum_{k=0}^n (-1)^{\frac{k(k-1)(k-2)(k-3)(k-4)(k-5)(k-6)(k-7)}{8!}} \la U \ra_k\\
&=&\sum_{k=0,1,\ldots, 7\mod 16}\la B\ra_k-\sum_{k=8, 9, \ldots, 15\mod 16}\la B\ra_k,\qquad \forall B\in\cl_{p,q}.
\end{eqnarray*}
In the general case, we have $(UV)^\square\neq U^\square V^\square$ and $(UV)^\square\neq V^\square U^\square$.
\end{ex}

Let us return to the case of arbitrary $n$. The scalar part operation $\la B \ra_0$ and the projection onto the center $\la B \ra_{\cen}$ can always be realized as linear combinations of operations of conjugation
$$B \mapsto \sum_{k=0}^n \lambda_k \la B \ra_k,\qquad \lambda_k=\pm 1.$$
This fact is discussed in \cite{det} (see Theorem 1 and Footnote 7). We need\footnote{Here and below we denote the integer part of $\log_2 n$ by $[\log_2 n$].}  $[\log_2 n]+1$ different operations of conjugation to do this\footnote{In the above example, we use the three operations $\widehat{B}$, $\widetilde{B}$, and $B^\va$ in the cases $4 \leq n \leq 7$, the four operations $\widehat{B}$, $\widetilde{B}$, $B^\va$, and $B^\square$ in the cases $8 \leq n \leq 15$.}.
 
Let us denote $m:=[\log_2 n]+1$. The formulas from Theorem \ref{thgen} and the following formulas\footnote{Note that $B^{\va_1}=\widehat{B}$, $B^{\va_2}=\widetilde{B}$, $B^{\va_3}=B^\va$, and $B^{\va_4}=B^\square$.} from \cite{det}
\begin{eqnarray*}
\la B \ra_0&=&\frac{1}{2^m} (B + B^{\va_1}+B^{\va_2}+\cdots +B^{\va_m}+B^{\va_1 \va_2}+\cdots +B^{\va_1 \ldots \va_m}),\\
B^{\va_j}&=&\sum_{k=0}^n (-1)^{C_k^{2^{j-1}}} \la B \ra_k,\qquad C_k^i=\frac{k!}{i! (k-i)!},\qquad j=1, \ldots, m,
\end{eqnarray*}
give us the basis-free solution (which involve only the operations of Clifford product, summation, and the operations of conjugation) to the Sylvester equation in $\cl_{p,q}$ with arbitrary $n$. Thus, our approach works in the case of arbitrary $n$.

\section{Conclusions}

In this paper, we present the basis-free solution to the Sylvester equation in Clifford (geometric) algebra of arbitrary dimension. Note that we discuss the most important (nondegenerate) case when the element $Q$ (\ref{Q1}), (\ref{Q2}), (\ref{Q3}), (\ref{Q4}), (\ref{Q5}), (\ref{Qn}), (\ref{Qnodd}) is non-zero and the corresponding Sylvester equation (\ref{Sylv123}) has a unique solution $X$. The degenerate case $Q=0$ (with zero divisors) can also be studied.

An interesting task is to generalize results of this paper to the case of general linear equations in geometric algebras
\begin{eqnarray}
\sum_{j=1}^k A_j X B_j=C,\qquad A_j, B_j, C, X\in\cl_{p,q}\label{gle}
\end{eqnarray}
in the case of arbitrary $n=p+q$. The basis-free solution to the equation (\ref{gle}) in the case of quaternions $\H\cong\cl_{0,2}$ is given in \cite{Hongbo} (see also the papers \cite{1,2}).

Note that all results of this paper remain true for the complexified Clifford algebras $\C\otimes\cl_{p,q}$ because characteristic polynomial coefficients are the same and are defined for $\cl_{p,q}$ and $\C\otimes\cl_{p,q}$ in the same manner using the matrix representation of $\C\otimes\cl_{p,q}$ of dimension $N=2^{[\frac{n+1}{2}]}$ (see \cite{det}).

The real Clifford algebras are isomorphic to the matrix algebras over $\R$, $\R\oplus\R$, $\C$, ${\mathbb H}$, or ${\mathbb H}\oplus{\mathbb H}$ depending on $p-q\mod 8$ and the complex Clifford algebras (and complexified Clifford algebras) are isomorphic to the matrix algebras over $\C$ or $\C\oplus\C$ depending on $n\mod 2$. In the opinion of the author, the structure of naturally defined fundamental subspaces (the subspaces of fixed grades, the even subalgebra, and the odd subspace) and the corresponding operations of conjugation (the grade involution, the reversion, and the others) favourably distinguishes Clifford algebras from the corresponding matrix algebras, when we use them for different applications -- in physics, engineering, robotics, computer vision, control theory, stability analysis, model reduction, image and signal processing.

The explicit formulas for solutions to the Sylvester and Lyapunov equations may be useful in applications, in particular, in solving algebraic noncommutative linear equations in quantum physics.

\subsection*{Acknowledgment}

This work is supported by the grant of the President of the Russian Federation (project MK-404.2020.1).

The research presented in this paper was stimulated by discussions of the author with Prof. Hongbo Li during scientific visit to the Chinese Academy of Sciences, Academy of Mathematics and Systems Science (Beijing, China) in 2019, for which the author is grateful. The author is grateful to Prof. Nikolay Marchuk for fruitful discussions.

The results of this paper were reported at the International Conference ``Computer Graphics International 2020 (CGI2020)'' (Geneva, Switzerland, October 2020). The author is grateful to the organizers and the participants of this conference for fruitful discussions.

The author is grateful to the editor, Prof. Eckhard Hitzer, and two anonymous reviewers for their careful reading of the paper and helpful comments on how to improve the presentation.

\end{document}